# Universal Lower Limit on Vortex Creep in Superconductors

S. Eley[1]*, M. Miura[2], B. Maiorov[1], L. Civale[1]

[1]Condensed Matter and Magnet Science, Los Alamos National Laboratory, Los Alamos, NM.

[2]Graduate School of Science & Technology, Seikei University, Tokyo, Japan.

*Correspondence to: seley@lanl.gov

**Superconductors are excellent testbeds for studying vortices, topological excitations that also appear in superfluids, liquid crystals, and Bose-Einstein condensates. Vortex motion can be disruptive; it can cause phase transitions[1], glitches in pulsars[2], and losses in superconducting microwave circuits[3], and it limits the current carrying capacity of superconductors[4]. Understanding vortex dynamics is therefore of fundamental and technological importance, and the competition between the effects of thermal energy and energy barriers defined by material disorder is not completely understood. In particular, early measurements of thermally-activated vortex motion (creep) in iron-based superconductors unveiled fast rates ($S$) comparable to measurements of YBa$_2$Cu$_3$O$_7$ (YBCO).[5–9] This was puzzling because $S$ is thought to somehow positively correlate with the Ginzburg number ($Gi$), and $Gi$ is orders of magnitude lower in most iron-based superconductors than in YBCO. Here, we report very slow creep in BaFe$_2$(As$_{0.67}$P$_{0.33}$)$_2$ films, and propose that there is a universal minimum realizable $S \sim Gi^{1/2}(T/T_c)$ (where $T_c$ is the superconducting transition temperature) that has been achieved in our films, few other materials, and violated by none. This limitation provides new clues on how to design materials**

**with slow creep and helps elucidate open questions regarding the interplay between system-specific length scales and vortex dynamics.**

The search for a universal description of vortex matter – one that is applicable to a range of systems and regimes – is a formidable challenge, complicated by the complexity of the interactions between vortices and the environment. Vortex motion that can be induced by Magnus and Lorentz forces or thermal activation can also be counteracted by pinning forces. Because vortex cores are normal (i.e., superfluidity or superconductivity is destroyed inside them), creating a vortex costs energy, and pinning can occur when it is energetically more favorable for a vortex to appear in one location than in another. In type-II superconductors at high enough magnetic fields, vortices are formed by the penetration of magnetic flux, and material disorder that locally reduces the vortex core energy can produce pinning forces that almost completely preclude vortex motion. This results in nearly zero resistance, as long as the current density ($J$) does not exceed the critical current density ($J_c$). The caveat is that, for $J<J_c$, vortices are in metastable states inside pinning centers that act as finite potential wells, and thermal fluctuations may enable them to jump out (creep). Creep is seldom problematic in applications involving low–$T_c$ superconductors (LTS), which typically exhibit slow creep (the decay over time in the persistent current, $S = \left|\frac{d\ln J}{d\ln t}\right|$). Consequently, Nb-Ti solenoids in MRI systems can operate in "persistent mode", retaining the field for essentially indefinite timeframes. However, creep is fast in high-$T_c$ superconductors (HTS), restricting applications and reducing the effective $J_c$.

Superconductors are ideal for studying fundamental questions regarding vortex dynamics because vortex motion introduces easily measureable dissipation into these materials. Furthermore, we can tune the strength of interactions between vortices and the potential energy landscape by incorporating pinning centers (e.g., secondary phases[4] such as BaZrO$_3$) or comparing materials with different properties (different *Gi*). The Ginzburg number, $Gi = (\gamma^2/2)\left[(\mu_0 k_B T_c)/(4\pi B_c^2(0)\xi_{ab}^3(0))\right]^2 \propto \gamma^2 T_c^2 \lambda^4/\xi^2$, parameterizes the scale of thermal fluctuations in a superconductor[10]; γ is the electronic mass anisotropy, $B_c(0) = \Phi_0/\left(2\sqrt{2}\pi\lambda_{ab}(0)\xi_{ab}(0)\right)$ is the thermodynamic critical field, $\Phi_0$ is the flux quantum, $\xi_{ab}(T=0)$ is the coherence length, and $\lambda_{ab}(T=0)$ is the penetration depth. As *Gi* in YBCO is significantly larger than in Nb-Ti (~$10^{-2}$ and $10^{-8}$, respectively), it is universally accepted that this difference roughly accounts for the much higher *S* in YBCO.[10] The same considerations suggest that iron-based superconductors (Fe-SCs) should have creep rates between those of LTS and HTS, but this is not what was observed. The exceptionally high *S* in many Fe-SCs[5–9] not only eludes conventional wisdom, but also stymies application. This problem highlights an important, yet unresolved, challenge: can we predict *S*(*T*) in any material?

To investigate the creep problem in Fe-SCs, we chose to study BaFe$_2$(As$_{0.67}$P$_{0.33}$)$_2$ because there is evidence of a strong pinning landscape in this material, demonstrated by high $J_c$[11] that can be further enhanced by introducing BaZrO$_3$ (BZO) nanoparticles[12] (Fig 1). In fact, the force exerted by the pinning landscape in films containing BZO rivals that in MgB$_2$, NbTi, and Nb$_3$Sn[12]. It has an intermediate $Gi \approx 8.7 \times 10^{-5}$, calculated using γ=1.55[12], $\lambda_{ab}(T=0) \approx 185$ nm (see Supplementary Information), and $\xi_{ab}(T=0) \approx \sqrt{\Phi_0/2\pi\mu_0 H_{c2}(0)} \approx 2.5$ nm [13]. Here, we compare results for two 80 nm-thick BaFe$_2$(As$_{0.67}$P$_{0.33}$)$_2$ films grown by pulsed laser deposition on MgO substrates[12], one containing a uniform dispersion of BZO nanoparticles (Ba122:P+BZO) and one without (Ba122:P).

Magnetization ($M \propto J_c$) studies were performed in a SQUID magnetometer (see Methods). By recording successive measurements of $M$ every 15s for an hour, at different temperatures ($T$) and fields ($H$), we capture the logarithmic decay over time in $J$ and determine both $J_c(T,H)$ and $S(T,H)$. We see that the nanoparticles increase $J_c$(5K, 200 Oe) by only ~12% and that this enhancement vanishes at high $T$ (Fig. 1b), but increases dramatically with increasing field (400% at 2T, Fig. 1c). This field-dependent enhancement in $J_c$ is not surprising. Similar to defects created by particle irradiation[4,14], the efficacy of the BZO in immobilizing vortices is magnified as the number of vortices approaches the number of available pinning sites. Consistent with other studies[11,12,15,16], $J_c$ is also high in films without nanoparticles. Strong pinning in this material has been ascribed to spatial variations in the P dopant concentration[11,15]. Namely, the $T_c$ and $\xi$ in BaFe$_2$(As$_{1-x}$P$_x$)$_2$ are exceptionally sensitive to the stoichiometry[16], such that these variations demarcate phase-separated regions that are effective pinning sites. These regions are spaced roughly 100 nm apart[15], therefore, outnumber the vortices in fields less than $B_\phi \sim 2\Phi_0/\sqrt{3}a_0^2 \approx 0.2 \text{ T}$. This is analogous to pinning in Sm$_{1+x}$Ba$_{2-x}$Cu$_3$O$_y$ (Sm123) films, in which Sm-rich123 precipitates of lower $T_c$ than the Sm123 matrix are effective pinning centers[17].

Fig. 2a shows $S(T)$ at different fields for both films, revealing behavior consistent with the Anderson-Kim (A.K.) model. A defect (or collection of defects) pins a vortex segment (or a bundle of vortices) by reducing the vortex line energy by the pinning energy, $U_p(T,H)$. The current-induced force reduces $U_p$ to an activation energy $U_{act}(T,H,J)$, so the vortex hopping rate is proportional to $e^{-U_{act}/T}$. All of the complexity of the creep phenomena is contained in $U_{act}(T,H,J)$. The A.K. model assumes $U_{act}(J) \propto U_P |1 - J/J_c|$, such that $S(T) \approx k_B T/U_P(T)$ [18]. For $T << T_c$, $U_p(T)$ is approximately constant and $S$ increases linearly with $T$; at higher temperatures, $S(T)$ steepens as $U_p(T)$ decreases. The data in Fig. 2a are well-described by this scenario, except that the curves extrapolate to nonzero values as $T$ extrapolates to zero, indicative of a measurable contribution from quantum creep (non-thermal, tunneling process)[10].

Comparing $S(T)$ in both films (Fig. 2a), we see that the addition of BZO increases $S$ for $\mu_0 H$ = 200 Oe and 0.3 T. Hence, BZO nanoparticles counteract the strong pinning effects of the large preexisting phase-separated regions. To understand this, note that in the Ba122:P+BZO film, vortices will pin to both the sparse phase-separated regions and the BZO. The BZO NPs are presumably smaller than the phase-separated regions, therefore, have lower $U_p$ such that vortices pinned to them will have higher $S$. At low fields (i.e., fewer vortices than phase-separated regions), we find overall higher $S$ for the Ba122:P+BZO film. The situation reverses at high fields. From Fig. 2a, we see that at $\mu_0 H \geq 1$ T, the film with nanoparticles shows slower $S(T)$ and, from Fig. 2b (inset), that this crossover occurs around $\mu_0 H$ = 0.5 T at $T$ = 4.5 K. So, above this field, the vortex population is high enough that BZO becomes effective in reducing $S$ as it offers more pinning sites to immobilize a higher fraction of vortices than the Ba122:P film.

From comparing our results with other studies, we find that $S$ in the Ba122:P film is not only remarkably slow, but is in fact the slowest measured $S$ in any Fe-SC to date (Fig. 3a). Prior to this study, we had speculated that $S(T)$ for other Fe-SCs seemed surprisingly high[5–8,19], which raised an important question: precisely what should we expect and how much lower can we go?

$Gi$ dictates various properties, including the width of the critical fluctuation region around $T_c$ (i.e., $\Delta T_c \sim Gi T_c$, setting the lower limit for the superconducting transition width for a crystallographically perfect sample), and the vortex melting transition[10,20]. It is known that $Gi$ will impact $S$ and has become perfunctory to mention an estimate of $Gi$ [6,21–32] when characterizing creep, but a testable formula has not been developed. In pursuit of such a formula, we now review how vortices depin from various pinning centers (applicable to crystals, films, devices, and wires). We restrict ourselves to the early stages of the depinning process ($J \sim J_c$), low $T$ (where $U_p$ is nearly temperature independent) and low $H$ [single vortex regime (sv)]. Vortex lines can be pinned either by the collective action of many weak pins or the independent action of strong pins[33]. First considering the most pervasive defect, point defects are weak pins that will only provide sufficient pinning when acting collectively. Here, $J_c \equiv J_{sv}$ is highest and the pinning energy[33,34][33,34] is given by[10] $U_P \equiv U_{sv} \sim H_c^2 \left( \xi_{ab}^3 / \gamma \right) \left( J_{sv} / J_0 \right)^{1/2}$, where $J_0$ is the depairing current density, the theoretical maximum achievable $J_c$. Consequently, the resulting creep rate is $S \sim T/U_P \sim \left( J_0 / J_{sv} \right)^{1/2} Gi^{1/2} \left( T/T_c \right)$, so the scale of $S$ depends on $Gi^{1/2}$. The prefactor $\left( J_0 / J_{sv} \right)^{1/2} > 1$ decreases as $J_{sv}$ increases, but in practice $J_{sv} \ll J_0$, thus $S > Gi^{1/2} (T/T_c)$.

Now consider a single vortex trapped by a strong individual pinning center, $U_P(T) \sim \left( H_c^2 / 8\pi \right) V_P$, where $V_p$ is the volume of the vortex core that is pinned. Strong pinning can occur in defects of dimensions $\sim \xi$ or larger. For an insulating inclusion of size equal to the coherence volume, i.e., an ellipsoid of semi axes $\xi_{ab}$ and $\xi_c = \xi_{ab}/\gamma$ such that $V_P \equiv V_{coh} \sim (4\pi/3)\xi_{ab}^3/\gamma$, we obtain $U_P \equiv U_{coh} \sim H_c^2 \xi_{ab}^3/6\gamma$, thus in the A.K. regime, $S \sim (T/U_P) \sim 6\sqrt{2} Gi^{1/2}(T/T_c)$, once again proportional to $Gi^{1/2}$.

The obvious strategy to minimize $S$ is to increase $U_p$ by increasing the length of the core that is pinned. For a defect of dimensions bigger than $\xi$ and arbitrary shape, $V_P \sim 2\pi \xi_{ab}^2 L_z \sim \frac{3}{2}(L_z/\xi_z)V_{coh}$, where $L_z > \xi_c$ is the size along the field direction (see Fig. 2c) and $\sqrt{2}\xi$ is the vortex core radius. There is, however, a limit to this approach set by the elastic properties of vortex matter, which are not considered by the A. K. model. This can be easily visualized by considering a vortex pinned in a columnar defect (CD), from which it can depin by producing a half-loop of length $\ell_{hl} \sim \xi_c (J_0/J)$ along the CD (Fig. 2c). Even for an arbitrarily long CD, the effective $L_z$ is truncated by $\ell_{hl}$ so $V_P \sim 2\pi \xi_{ab}^2 \ell_{hl}$. For $J \sim J_c$, the activation energy $U_{act}(J) \sim (4\pi/\gamma)H_c^2 \xi_{ab}^3 (J_0/J_c)(1 - J/J_c)$ [35] resulting in

$$S \sim 4\sqrt{2}(J_c/J_0)Gi^{1/2}(T/T_c).$$

For randomly distributed NPs of radius $R \gg \xi(T=0)$, with increasing $J$ the tips of the pinned vortex slide along the surface of the NP until they meet near the NP equator and reconnect, leading to depinning of the vortex (Fig. 2c)[36]. As $J$ decays during relaxation, the two tips progressively separate. While $J \sim J_c$, the separation is small and a thermal fluctuation that creates a short vortex segment that connects the tips is enough to depin the vortex, thus $U_{act}$ is small and $S$ is large regardless of $R$. The length of the segment required for a depinning fluctuation increases with decreasing $J$, resembling the case of a CD where $L_z \sim 2R$. If $R$ is very large and the energy of creating a segment connecting both tips becomes too large, depinning may occur through the creation of a localized excitation analogous to a CD half-loop, and the above estimate for $S$ is recovered. The argument can be extended to large defects of arbitrary shape and, while the numerical details may be system-dependent, a relation of the form $S \sim A*Gi^{1/2}(T/T_c)$, where $A$ is an unknown proportionality constant, will still hold.

To estimate $A$, we start with $A \sim 4\sqrt{2}(J_c/J_0)$ for a CD. In systems with vortex-dominated dissipation, a maximum of $J_c/J_0 \sim 0.2$-$0.25$ [37] has been achieved, so $A \sim 1$. For defects of diverse morphology, we obtain an estimate for $A$ empirically using a wealth of $S(T)$ data from YBCO samples with different microstructures (Fig. 3c) and see that again $A \sim 1$. Thus, we conclude that, at low temperature (where $U_p$ is nearly temperature-independent) and low field (where vortex-vortex interactions are negligible), $S(T)$ cannot be significantly smaller than $S \sim Gi^{1/2}(T/T_c)$ for $J \sim J_c$ (see Supplementary Information for further discussion). This reveals a fundamental limitation to how much creep can be slowed through modifications of the microstructure, and can serve as a guide for when further improvements can be achieved. In light of this, we find that $S$ in our films has the same relationship to $Gi$ as we found in the YBCO films with the lowest $S$ values, and that our films have reached the slowest $S$ allowed by $Gi$.

By compiling results from the literature for different superconductors, we reveal a cogent picture of this limitation and find few materials, including our Ba122:P film, MgB$_2$[38,39], and YBCO, with measured creeps rates near our proposed limit (Fig. 3b). Fig. 4 shows $S(T=T_c/4, \mu_0H=1$ T$)$ for many SC materials covering a broad range of values for $Gi$.

Strikingly, the straight line at $S \sim Gi^{1/2}/4$ demarcates a region of unobtainable *S* values; no data appears below it. This hard constraint, which had not been established until now, has two broad implications: first, the creep problem in HTS cannot be fully eliminated and there is a limit to how much it can be ameliorated, and secondly, we can confidently predict that any yet-to-be-discovered HTS will have fast creep. Additionally, these results will help in designing materials for superconducting devices, wires, and magnets, and could provide insight on open questions in other systems in which topological excitations undergo depinning processes, e.g., skyrmions in magnetic films[40].

**Figures.**

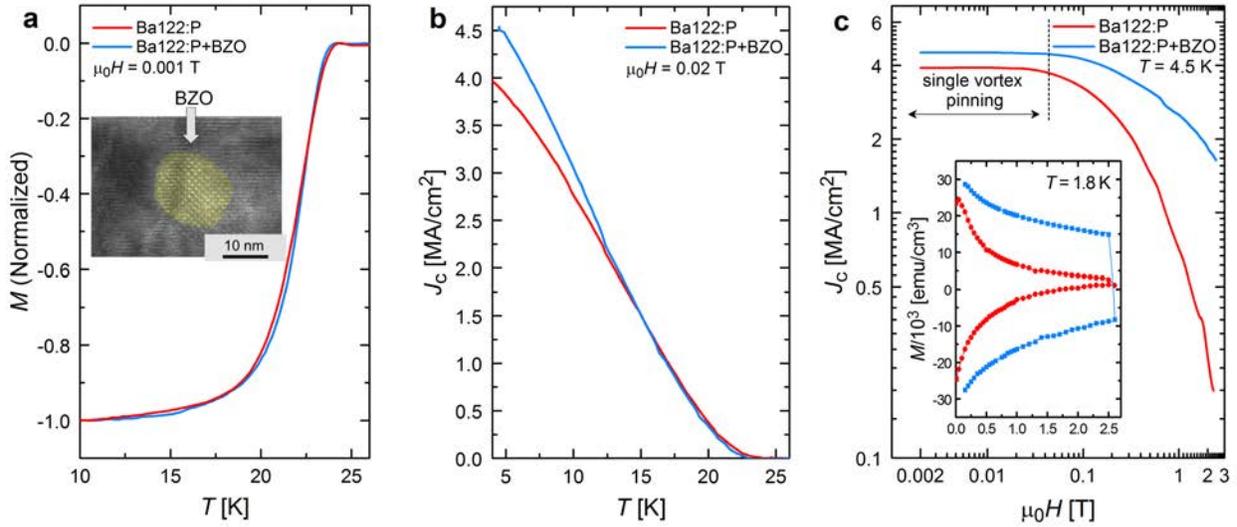

**Figure 1| Enhancement in the Critical Current of BaFe$_2$(As$_{0.67}$P$_{0.33}$)$_2$ films with the addition of BaZrO$_3$ (BZO) nanoparticles.** Comparison of the temperature dependence of the (a) magnetization (*M*) at $\mu_0H$=0.001 T and (b) critical current density (*J*$_c$) at $\mu_0H$=0.02 T in Ba122:P and Ba122:P+BZO. The BZO NPs are ~8 nm in diameter and spaced 24 nm apart (i.e., density of 6.8×10$^{22}$ m$^{-3}$). The addition of nanoparticles does not degrade the critical temperature, $T_c \approx 24.7$ K. The inset to (a) is a cross-sectional high-resolution TEM image of the Ba122:P+BZO film, in which a BZO nanoparticle can be easily discerned. (c) Field dependence of *J*$_c$ at *T*=4.5 K and (inset) magnetization at *T*=1.8 K.

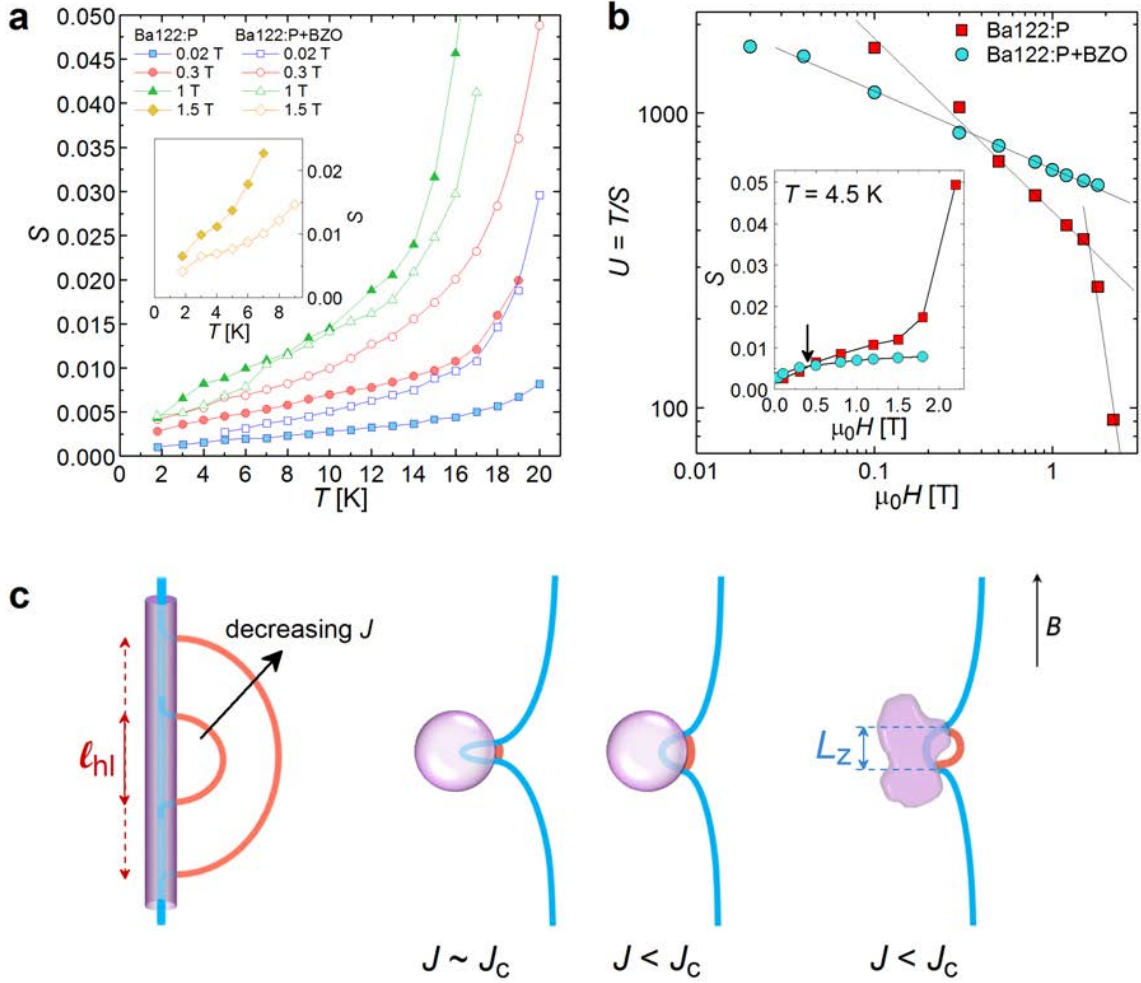

**Figure 2| Temperature and field dependence of creep.** **(a)** Comparison of creep, $S(T)$, in Ba122:P and Ba122:P+BZO at $\mu_0 H = 0.02$ T, 0.3 T, 1 T and (inset) 1.5 T. **(b)** Comparison of field-dependent activation energies, $U_P = T/S$, and (inset) creep at $T$=4.5 K in both samples. The BZO nanoparticles become effective in lowering creep in high fields $\mu_0 H \geq 0.5$ T. **(c)** Illustration of excitations of a vortex line from a columnar defect, spherical nanoparticle (in the early, then later, stages of relaxation), and nanoparticle of arbitrary shape through half-loop formation, leading to creep. The size of the loop grows with decreasing current density $J$.

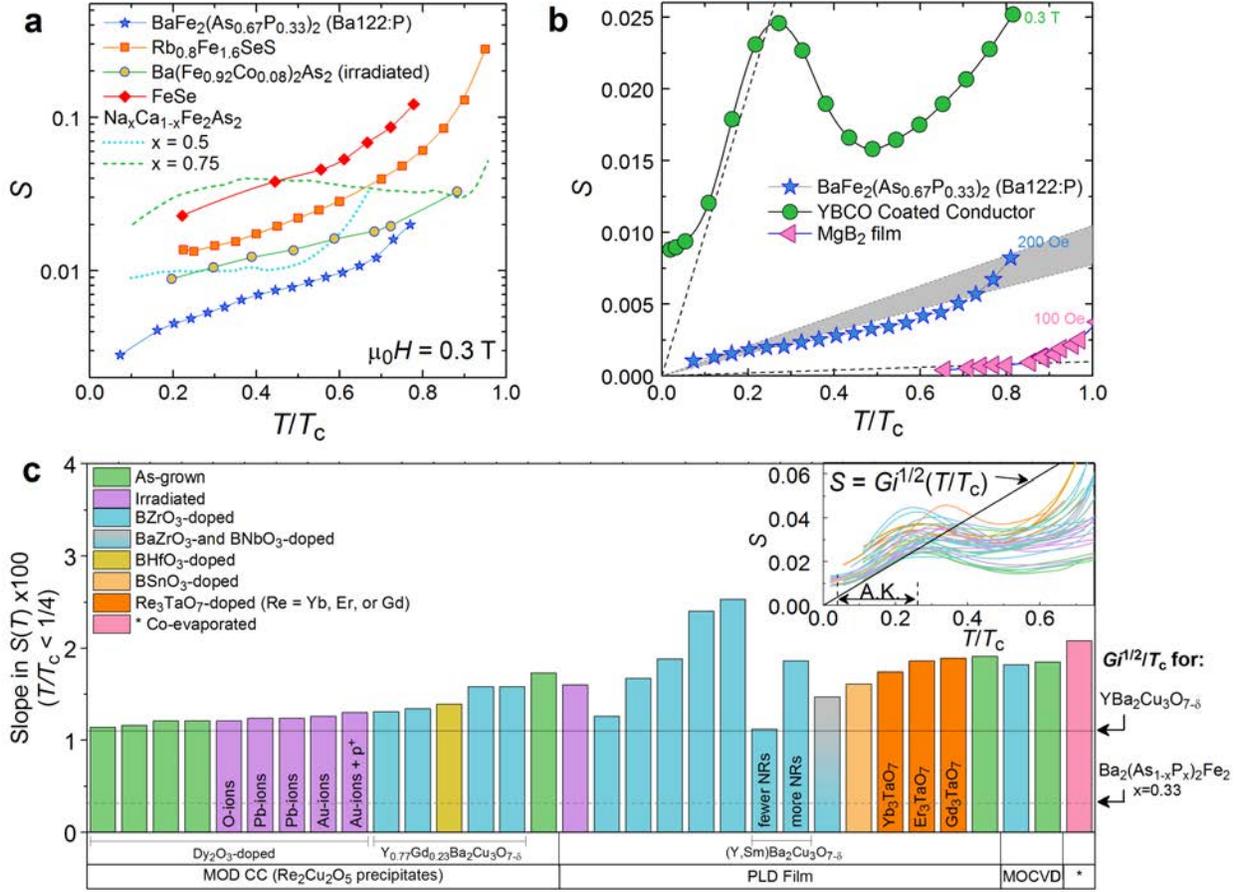

**Figure 3| Creep in Iron-based Superconductors, MgB$_2$, and YBa$_2$Cu$_3$O$_{7-\delta}$. (a)** A comparison of creep vs reduced temperature, $S(T/T_c)$, in different iron-based superconductors (Fe-SCs)[8,23], where the data for FeSe was extracted from $S(H)$ at different temperatures[9] and we measured Rb$_{0.8}$Fe$_{1.6}$SeS [41]. All previous studies of creep in Fe-SCs have been performed on single crystals[5–8,14,19,22,23,32,42–44], where self-fields ($H_{sf} \propto J_c \times$ thickness) are higher than in our films. Consequently, we perform a direct comparison at $\mu_0 H = 0.3$ T because it is higher than self-field in all samples, is a relatively low field which typically results in lower $S(T)$ than higher fields, and was a common choice in available data. **(b)** $S(T/T_c)$ at the indicated fields (lowest field used that is above self-field) in Ba122:P, a YBa$_2$Cu$_3$O$_{7-\delta}$ coated conductor[45], and MgB$_2$[39]. The limit set by $Gi^{1/2}T/T_c$ is shown

for each respective film by the dashed lines. The gray region captures the uncertainty in $Gi^{1/2}T/T_c$, due to the uncertainty in λ (170 nm from μSr measurements and 200 nm from tunnel-diode oscillator[13]) for Ba122:P. **(c)** The histogram shows the slope in *S* vs *T* at $\mu_0H$ = 0.3 T, for YBCO (or, if specified, $Y_{1-x}Re_xBCO$ where Re=Gd or Sm) samples with a wide range of pinning landscapes (see Supplementary Information). The inset shows the raw data. The slope was extracted for $T/T_c \leq 1/4$ (Anderson-Kim regime for these samples) and quantum creep is not considered (the fit is forced through $T = 0$, $S = 0$).

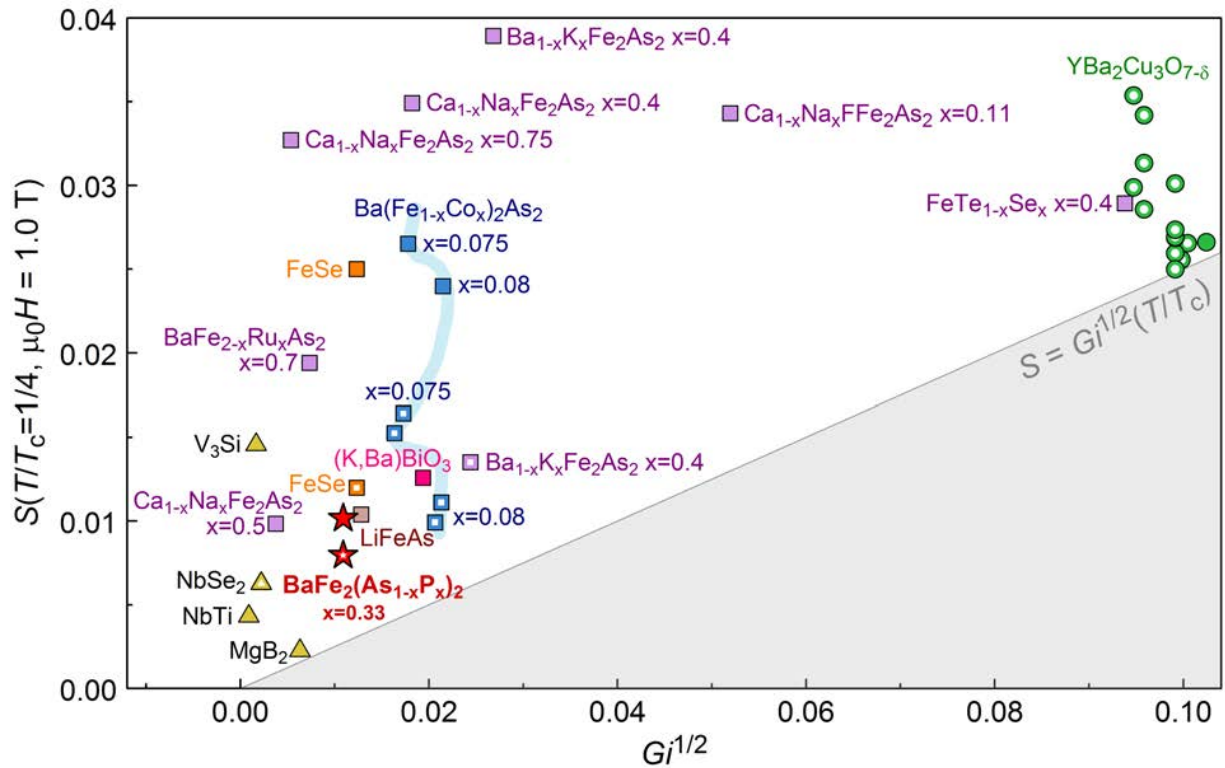

**Figure 4| The universal lower limit for creep. (A)** Creep at reduced temperature $T/T_c=1/4$ and field of $\mu_0H$ = 1 T for different superconductors (see Supplementary Information) plotted versus $Gi^{1/2}$. The open symbols indicate materials for which the microstructure has been modified by either irradiation or incorporation of inclusions. The solid gray line represents the limit set by $Gi^{1/2}T/T_c= Gi^{1/2}/4$. We have chosen $T=T_c/4$

because it is a low temperature where the A.K. description is typically valid (see Supplementary Information), and $\mu_0 H$=1 T because it is the lowest field for which a large amount of $S$ data is available in the literature.

**Acknowledgements.**

This work was funded by the U.S. DOE, Office of Basic Energy Sciences, Materials Sciences and Engineering Division (S.E., B.M., and L.C.). Sample fabrication was supported by the Japan Society for the Promotion of Science through the "Funding Program for World-Leading Innovation R&D on Science and Technology."  M. M. was supported by JSPS KAKENHI (26709076).

**Author Contributions.**

S.E. carried out the magnetization measurements and data analysis. L. C. and B.M designed the experiment, and L.C. carried out the theoretical analysis. M.M. grew and performed microstructural characterization on the films.  All authors contributed to manuscript preparation.

**Competing Financial Interest Statement.**

The authors declare no competing financial interests.

## Methods

**Film Growth and Characterization.**

We used the film growth procedures reported previously[12]. Epitaxial films (80 nm-thick) were deposited on MgO (100) single crystalline substrates, held at 800°C, by ablating pulsed laser deposition (PLD) targets using the second harmonic (wavelength: 532 nm) of a pulsed Nd:YAG laser at a repetition rate of 10Hz in vacuum of $10^{-4}$ Pa. Targets of nominal composition $BaFe_2(As_{0.67}P_{0.33})_2$ and $BaFe_2(As_{0.67}P_{0.33})_2$+3mol.% $BaZrO_3$ (BZO) were made by sintering Ba metal (99.99%, chunk), Fe metal (99.9%, no. 300 mesh pass), As powder (99.9999%, B3 mm grain) and P powder (99.999%, no. 600 mesh pass). Mixtures of 1:1 Ba:As and $FeAs_{0.17}P_{0.33}$ were each separately heat treated in evacuated quartz tubes at 500°C for 10 hours and 650°C for 20 hours. The products were powdered and reacted at 650°C for 10 hours and 900°C for 24 hours. $BaZrO_3$ powder (3 mol%) was added to the resulting $BaFe_2(As_{0.67}P_{0.33})_2$ powder and reacted at 1,000°C for 24 hours to prepare the target with $BaZrO_3$. Transmission Electron Microscopy (TEM) was used to analyze the microstructure and measure the film thickness. Energy dispersive x-ray spectroscopy and high-resolution TEM revealed stacking faults around the BZO inclusions[12].

**Magnetization Measurements.**

Magnetization studies were performed using a Quantum Design SQUID magnetometer to determine the critical temperature, $T_c$, and characterize the temperature and field dependence of $J_c$ and $S$. The magnetization ($M$) was obtained by dividing the measured magnetic moment ($m$) by the sample volume $V = \delta w l$, where $\delta$ = 80 nm is the film thickness, and $w \approx l \approx 3.5$ mm specifies sample width and length. Measurements of the Meissner slope confirmed that the superconducting volume was equivalent to sample dimensions. For all measurements, the magnetic field was applied parallel to the c-axis (perpendicular to the film plane). $T_c$ was determined from magnetization as a function of temperature at $\mu_0 H$ = 0.001 T. The critical current was calculated using the Bean critical-state model [46,47], $J_c = 20\Delta M/\left[w\left(1-\frac{w}{3l}\right)\right]$, where $\Delta M$ is the difference between the upper and lower branches of isothermal magnetization loops $M(H)$.

Creep data were taken by using standard methods [18] for measuring the time-dependent magnetization, $M(t)$. First, the critical state was established by sweeping the field $\Delta H > 4H^*$, where $H^*$ is the minimum field at which magnetic flux will fully penetrate the sample. Magnetization is then recorded every ~15 s; a brief measurement in the lower branch is collected to determine the background, then the upper branch is measured as a function of time ($t$) for an hour fixed at the specified field. After subtracting the background and adjusting the time to account for the difference between the initial application of the field and the first measurement (maximize correlation coefficient), $S = -d \ln M/d \ln t$ is extracted from the slope of a linear fit to ln $M$-ln $t$.

# Supplementary Information

# Universal Lower Limit on Vortex Creep in Superconductors

S. Eley[1]*, M. Miura[2], B. Maiorov[1], L. Civale[1]

[1]Condensed Matter and Magnet Science, Los Alamos National Laboratory, Los Alamos, NM.

[2]Graduate School of Science & Technology, Seikei University, Tokyo, Japan.

*Correspondence to: seley@lanl.gov

**Calculation of the Dependence of *S* on *Gi*.**

The ideal superconductor for applications should have the largest possible $J_c$ and lowest possible *S*. Our claim that *S* cannot be significantly lower than $S \sim Gi^{1/2}(T/T_c)$ is restricted to the Anderson-Kim (A.K.) regime and pertains to the best ratio $J_c/J_0 \sim 0.2 - 0.25$ that has so far been achieved in real superconducting wires. Models predict that an ideal CD should produce $J_c \sim J_0$; in that case $\ell_{hl} \sim \xi_c$ and $S \sim 4\sqrt{2}\left(\frac{J_c}{J_0}\right) Gi^{1/2}(T/T_c)$ should be 4-6 times higher. Generalizing from the case for a CD $\sim 4\sqrt{2}\left(\frac{J_c}{J_0}\right) Gi^{1/2}(T/T_c)$, and acknowledging that the numerical factors may be dependent on the pinning landscape, we expect that, if higher $J_c$s are attained in the future, the creep rates will also increase. Moving in the opposite direction, much lower creep rates (data points in the "forbidden region" of Fig. 4) could be obtained for systems where $J_c/J_0 \ll 0.2$, but those are not of interest for applications.

Now, let us discuss what happens outside the limits of our claim. In the simple example discussed in the main text of a single vortex trapped by a strong individual pinning center, where $U_p(T) \sim \left(\frac{H_c^2}{8\pi}\right) V_p$, as *T* increases $H_c(T)$ and $U_p(T)$ decrease, so *S* at high *T* may increase faster than linearly with *T* while still remaining in the A.K. regime, as may be the case in Fig. 2a and in the Ba122:P and MgB$_2$ data in Fig. 3b. In materials with fast creep, where the condition $J \ll J_c$ may be reached at high *T* or in later stages of the relaxation process, the situation in the case of large defects or collective pinning is more complex. We can again use the case of the depinning process from CD via half-loops as an example. As *J* decreases, $\ell_{hl}$ grows and diverge in the limit $J \to 0$, in contrast to the Anderson-Kim (A.K.) scenario. The activation energy is given by the interpolation formula

$U_{act}(J) \sim H_c^2 \xi_{ab}^2 \left(\frac{J_0}{J_c}\right)\left(\frac{J_c}{J} - 1\right)$ and also diverges for $J \to 0$. This is a manifestation of glassiness in the vortex dynamics, which arises from the fact that vortices are extended elastic objects rather than point particles and implies that $S \to 0$ for $J \to 0$. Glassiness also occurs in the case of collective pinning by random point defects[10]. Depending on the details of the pinning mechanism, $S(T)$ may exhibit a plateau[48] or even decrease with *T*, as in the YBa$_2$Cu$_3$O$_{7-\delta}$ (YBCO) data in Fig. 3b. Glassy vortex relaxation is conceptually important, leading to the notion of a "true superconductor", and has been the subject of extensive research, but it's quite irrelevant for applications and for our analysis. We are interested in the early stages of the relaxation process, where *J* is close to $J_c$.

**Comparison of YBa$_2$Cu$_3$O$_{7-\delta}$ Samples.**

The histogram in Fig. 3c in the main text shows the slope of $S(T/T_c \leq 1/4)$ for many different YBCO samples. Most of the samples have additions of second phases (e.g., BaZrO$_3$ nanoparticles), rare-earth additions or substitutions (namely, mixtures of Y and a rare-earth element), or have undergone irradiation to test different pinning enhancement mechanisms with the goal of increasing $J_c$ [4]. The data includes YBCO and (Y,Gd)BCO coated conductors (CCs) grown by metal organic deposition (MOD)[45,49-52]. It also includes a variety of films grown by pulsed laser deposition (PLD): in one study, different growth temperatures were used to modify the orientation and length of BZO inclusions[53], others incorporated BaZrO$_3$ (BZO) and BaNbO$_3$ inclusions[54] or Re$_3$TaO$_7$ nanoparticles (where Re=Yb, Er, or Gd)[54,55]. Additionally, (Y,Sm)BCO films were grown with two different densities of BZO nanorods[54] using PLD. Lastly, the diagram includes two films grown using metal organic chemical vapor deposition (MOCVD) and one using co-evaporation[54]. See Supplementary Table S1 for more information.

**Parameters for Ginzburg Number and Creep Rate Comparisons.**

Fig. 4 in the main text is the result of a fairly exhaustive literature search for creep rates measured under comparable conditions for a range of superconductors. We attempted to choose parameters (μ$_0$H=1 T, $T/T_c$=1/4) for which we could find the widest selection of data within the Anderson-Kim regime at fields above self-field. (Note that we have repeated the analysis at μ$_0$H=0.3 T and $T/T_c$=1/4; the results are shown in Fig. S1.) It not only reveals a trend, but also a gap within the range $Gi \sim$ 1-8x10$^{-3}$ in which it would be interesting to study more materials. The material parameters used to calculate the Ginzburg numbers, and corresponding references, for the data plotted in Fig. 4 in the main text are tabulated in Table S2. Unless otherwise specified (with a reference), the coherence length was calculated from $\xi_{ab}(0) = \sqrt{\Phi_0/2\pi\mu_0 H_{c2}(0)}$. Supplementary Table S3 specifies the corresponding measured creep rates.

- Co-doped 122

We were unfortunately unable to include $S$ measured for Ba(Fe$_{0.93}$Co$_{0.07}$)$_2$As$_2$ from Ref. [14] in Fig. 4 from the main text because the data at $\mu_0 H$= 1.0 T was measured at $T/T_c \approx 1/5$. Two crystals, one irradiated with 2.6 GeV U-ions and the other with 200

MeV Au-ions, both showed $S \approx 0.0087$ whereas $Gi^{1/2}/5 \approx 0.0025$. Furthermore, we have included $S$ measured in Ref. [23] for single crystals of $Ba(Fe_{0.92}Co_{0.08})_2As_2$ after annealing, then after undergoing irradiation at two different doses. It is possible that the sample irradiated with the highest dose has $S$ values at or near our proposed limit at lower fields than used in the reported measurements.

- $YBa_2Cu_3O_{7-\delta}$ Samples

    For Fig. 4 in the main text, we chose to compare $S$ in different materials at $\mu_o H = 1T$, because it was the lowest field for which there was a wide range of data in the literature, and $T/T_c = 1/4$ because it is typically within the Anderson-Kim regime. It should be noted, however, that in $YBa_2Cu_3O_{7-\delta}$ samples at 1 T, $T/T_c=1/4$ is slightly beyond the Anderson-Kim regime, where $S$ is lower than would be predicted by $S \sim T/U_P$ due to glassy effects or strong pinning from nanoparticles. For calculation of $Gi$ in YBCO samples, we use $\xi_{ab}(0) = 1.6$ nm, $\lambda_{ab}(0) = 140$ nm, $\gamma = 5$ [10], and the measured $T_c$, resulting in $Gi \sim 9 \times 10^{-3} - 0.01$.

**Uncertainty in the Ginzburg Number.**

The predominate sources of uncertainty in our $Gi$ calculations will come from $\lambda$ and $\gamma$. Penetration depth measurements often have uncertainties of ~20-70 nm[56,57], depending on the measurement technique. Additionally, how irradiation affects $\lambda$ has not been well-studied, though there is a study[58] that captures increases in $\lambda$ in YBCO single crystals in which $T_c$ has been significantly suppressed by irradiation. For mulitband superconductors, the anisotropy of the critical field is temperature dependent, typically near 1 at low temperatures and higher near $T_c$. It is not obvious which value is most appropriate for our calculation. We chose to use $\gamma(T/T_c=0.9)$ because it represents the real electron mass anisotropy, whereas the measured ratio of $H_{c2}$ || ab and $H_{c2}$ || c at $T\rightarrow 0$ can be affected by pair breaking. Also, note that our calculation may differ from *Gi* values mentioned in other references. Sometimes the inference that the penetration depth ($\lambda$) in materials of the same class but different stoichiometry is the same is incorrect, and results in inaccurate estimates of *Gi*. (Note that we also use estimated values for $\lambda$ based on results for samples with similar stoichiometry in the case of three materials plotted in Fig 4 in the main text for which $\lambda$ has not been measured. See Supplementary Table S2.)

**Supplementary Table S1. The measured creep rates, and corresponding references, for YBCO samples in Figures 3c and 4 of the main text.** If applicable, the method of enhancing $J_c$ by irradiation (including dose in terms of flux or matching field) or integration of precipitates/impurities is noted. For simplicity, the following abbreviations are used: PLD (pulse laser deposition), MOD (metal organic deposition), CDs (columnar defects), NPs (nanoparticles), and NRs (nanorods).

| $J_c$ Enhancement Method (Inclusions or Irradiation) | S 0.3 T | 1.0 T | Ref for S |
|---|---|---|---|
| **Single Crystal, Flux growth-method** | | | |
| Pristine | | 0.02545 | 59 |
| Irradiated, 3 MeV H+, $1 \times 10^{16}/cm^2$ | | 0.02058 | 59 |
| **MOD YBCO Coated Conductors** | | | |
| $Y_2Cu_2O_5$, $Dy_2O_3$ | 0.02407 | 0.02656 | |
| $Y_2Cu_2O_5$, $Dy_2O_3$ | 0.02407 | 0.02557 | 45 |
| $Y_2Cu_2O_5$, $Dy_2O_3$ | 0.02559 | 0.02596 | |
| $Y_2Cu_2O_5$, $Dy_2O_3$ | 0.02446 | 0.02501 | |
| $Y_2Cu_2O_5$, $Dy_2O_3$, 6 MeV O-ion, $5 \times 10^{13}/cm^2$ | 0.02614 | 0.02989 | 45 |
| $Y_2Cu_2O_5$, $Dy_2O_3$, 1.4 GeV Pb-ion, 3T (CDs parallel to c-axis) | 0.02821 | 0.03538 | |
| $Y_2Cu_2O_5$, $Dy_2O_3$, 1.4 GeV Pb-ion, 3T (CDs 30° from c-axis) | 0.02856 | 0.04177 | |
| $Y_2Cu_2O_5$, $Dy_2O_3$, 18 MeV Au-ion, $6 \times 10^{11}/cm^2$ | 0.02645 | 0.03134 | 50 |
| $Y_2Cu_2O_5$, $Dy_2O_3$, 18 MeV Au-ion and H+, $6 \times 10^{11}/cm^2$ | 0.02953 | 0.03420 | |
| $Y_2Cu_2O_5$, $Dy_2O_3$, 4 MeV H+, $8 \times 10^{16}/cm^2$ | | 0.02858 | 45 |
| $Y_2Cu_2O_5$ | 0.02981 | | 51 |
| **MOD YGdBCO Coated Conductors** | | | |
| $BaZrO_3$ | 0.02775 | 0.02688 | |
| $BaZrO_3$ | 0.02963 | 0.02692 | |
| $BHfO_3$ | 0.03060 | 0.03012 | |
| $BaZrO_3$ | 0.02897 | | 51 |
| $BaZrO_3$ | 0.02610 | 0.02736 | |
| **PLD** | | | |
| No inclusions | 0.03073 | | 54 |
| $BaZrO_3$, annealed at 745°C | 0.02367 | | 53 |
| $BaZrO_3$, annealed at 800°C | 0.03165 | | 53 |
| $BaZrO_3$, annealed at 815°C | 0.03431 | | 53 |
| $BaZrO_3$, annealed at 830°C | 0.04288 | | 53 |
| $BaZrO_3$, annealed at 840°C | 0.04495 | | 53 |
| $BaZrO_3$ (Y,Sm)BCO, NRs | 0.02205 | | 54 |
| $BaZrO_3$, (Y,Sm)BCO, NRs (higher density than sample on previous line) | 0.03240 | | 54 |
| $BaZrO_3$ and $BNbO_3$ | 0.03380 | | 54 |
| $BaSnO_3$ | 0.03611 | | 54 |
| $Yb_3TaO_7$ | 0.03458 | | 54 |
| $Er_3TaO_7$ | 0.03720 | | 54 |
| $Gd_3TaO_7$ | 0.03709 | | 54 |
| Heavy ion irradiated | 0.03408 | | 54 |
| **MOCVD** | | | |
| No inclusions | 0.03664 | | 54 |
| $BaZrO_3$ | 0.03862 | | 54 |
| **Co-evaporated** | | | |
| No inclusions | 0.03352 | | 54 |

**Supplementary Table S2. Parameters, and corresponding references, used to calculate the Ginzburg number for each material plotted in Figure 4 in the main text.** If applicable, the method of enhancing $J_c$ by irradiation (including dose in terms of flux in units of ions/cm$^2$ or matching field) or growth of precipitates/impurities is noted. The following abbreviations are used: irr (irradiated), CDs (columnar defects), and NPs (nanoparticles).

| Material | Type, Film deposition process, $J_c$ Enhancement Method | $T_c$ [K] | $H_{c2}(0)$ | $\xi_{ab}$ [nm] | $\lambda_{ab}$ [nm] | $\gamma_H$ ($T/T_c \approx 0.9$) |
|---|---|---|---|---|---|---|
| **LTS** | | | | | | |
| Nb | Foil | 9.3 | 0.5 | 24.8 | 80 [60] | 1 |
| NbTi | Wire | 9.3 | 14 [61] | 4.9 | 160 [61] | 1 |
| NbSe$_2$ | Crystal | 7 | 4.5 [62] | 8.6 | 250 [62] | 2.4 [63] |
| NbSe$_2$ | Crystal, Irr: 1.4 GeV Pb-ion (CDs*) | 7 | 4.5 [62] | 8.6 | 250 [62] | 2.4 [63] |
| **A15, MgB$_2$, BKBO** | | | | | | |
| V$_3$Si | Crystal | 17 [64] | 34 [64] | 3.1 | 130 [65] | 1 |
| MgB$_2$ | Film | 39.7 [39] | 3.5 [39] | 9.7 | 50 [39] | 2.5 [39] |
| MgB$_2$ | Film | 39.1 [38] | 13.2 | 5 [38] | 100 [38] | 2.5 [39] |
| (K,Ba)BiO$_3$ | Crystal | 30.8 [66] | 30 [66] | 3.3 | 340 [67] | 1 [68] |
| **Borocarbides** | | | | | | |
| LuNi$_2$B$_2$C | Crystal | 15.7 [69] | 7 [70] | 6.9 | 275 [71] | 1 [72] |
| **Fe-SCs: 122 P-doped** | | | | | | |
| Ba(As$_{0.67}$P$_{0.33}$)$_2$Fe$_2$ | Film | 24.7 | 51.1 [13] | 2.5 | 200 [13] | 1.55 [12] |
| Ba(As$_{0.67}$P$_{0.33}$)$_2$Fe$_2$ | Film, 3% BaZrO$_3$ NPs | | | | | |
| **Fe-SCs: 122 Na-doped** | | | | | | |
| Ca$_{0.5}$Na$_{0.5}$Fe$_2$As$_2$ | Crystal | 19.4 [8] | 24 | 3.7 [22]† | 146 [22] | 1.85 [22] |
| Ca$_{0.25}$Na$_{0.75}$Fe$_2$As$_2$ | Crystal | 33.4 [8] | 76 | 2.1 [22]† | 100 [22] | 1.85 [22] |
| Ca$_{0.6}$Na$_{0.4}$Fe$_2$As$_2$ | Crystal | 35 [7] | 48.5 [7] | 2.6 | 194 [73]‡ | 2 [7] |
| Ca$_{0.89}$Na$_{0.11}$FFeAs | Crystal | 34.5 [7] | 71.6 [7] | 2.1 | 190‡ | 5 [7] |
| **Fe-SCs: 122 Co-doped** | | | | | | |
| Ba(Fe$_{0.92}$Co$_{0.08}$)$_2$As$_2$ | Crystal§ | 26 [23] | 50 [74] | 2.6 | | |
| Ba(Fe$_{0.92}$Co$_{0.08}$)$_2$As$_2$ | Crystal§, Irr: 1.4 GeV Pb-ion (1 T) | 25.8 [23] | 50 [74] | 2.6 | | |
| Ba(Fe$_{0.92}$Co$_{0.08}$)$_2$As$_2$ | Crystal§, Irr: 1.4 GeV Pb-ion (3.5 T) | 25 [23] | 50 [74] | 2.6 | 260 [57,75] | 1.74 [76] |
| Ba(Fe$_{0.925}$Co$_{0.075}$)$_2$As$_2$ | Crystal | 24.4 [32] | 39 [74] | 2.9 | | |
| Ba(Fe$_{0.925}$Co$_{0.075}$)$_2$As$_2$ | Crystal, Irr: 3 MeV H+, 1x10$^{16}$/cm$^2$ | 23.7 [32] | 39 [74] | 2.9 | | |
| Ba(Fe$_{0.925}$Co$_{0.075}$)$_2$As$_2$ | Crystal, Irr: 3 MeV H+, 2x10$^{16}$/cm$^2$ | 22.4 [32] | 39 [74] | 2.9 | | |
| **Fe-SCs: Other 122** | | | | | | |
| BaFe$_{1.3}$Ru$_{0.7}$As$_2$ | Crystal | 18 [6] | 22 | 3.9 [6] | 210 [6] | 2 [6] |
| Ba$_{0.6}$K$_{0.4}$Fe$_2$As$_2$ | Crystal | 38.6 [42] | 94 [77] | 1.9 | 190 [78] | 2 [77] |
| Ba$_{0.6}$K$_{0.4}$Fe$_2$As$_2$ | Crystal, Irr: 3 MeV H+, 5.8x10$^{16}$/cm$^2$ | 35.1 [42] | 94 [77] | 1.9 | 190 [78] | 2 [77] |
| **Fe-SCs: Misc** | | | | | | |
| FeSe | Crystal | 9 [19] | 15 [79] | 4.7 | 445 [80] | 1.8 [79] |
| FeSe | Crystal, Irr: 3 MeV H+, 5x10$^{16}$/cm$^2$ | 9 [19] | 15 [79] | 4.7 | 445 [80] | 1.8 [79] |
| FeTe$_{0.6}$Se$_{0.4}$ | Crystal | 14.5 [5] | 48 [81] | 2.6 | 560 [82] | 3 [81] |
| LiFeAs | Crystal | 16.5 [44] | 15 [83] | 4.7 | 210 [84] | 2.4 [83] |
| Rb$_{0.8}$Fe$_{1.6}$SeS | Crystal | 19.5 | | | | |
| **LSCO** | | | | | | |
| La$_{1.85}$Sr$_{0.15}$CuO$_4$ | Crystal | 34 | 82 [85] | 2.0 | 240 [86] | 13 [61] |

---

* CDs are at a 30° angle with respect to the c-axis.

† The reference accidentally mixes up $\xi_{ab}$ and $\xi_c$.

‡ Has not been measured. We used value measured in similar material.

§ Annealed

**Supplementary Table S3. The measured creep rates, and corresponding references, for the data plotted in Figure 4 in the main text.** If applicable, the method of enhancing $J_c$ by irradiation (including dose in terms of flux in units of ions/cm$^2$ or matching field) or growth of precipitates/impurities is noted. The following abbreviations are used: irr (irradiated), CDs (columnar defects), and NPs (nanoparticles).

| Material | Type, Film deposition process, $J_c$ Enhancement Method | $G_i$ | $S$ 0.3 T | $S$ 1.0 T | Ref for S |
|---|---|---|---|---|---|
| **LTS** | | | | | |
| Nb | Foil | 1.9x10$^{-9}$ | 6.33x10$^{-4}$ | — | |
| NbTi | Wire | 7.9x10$^{-7}$ | 0.00235 | 0.00432 | |
| NbSe$_2$ | Crystal | 4.9x10$^{-6}$ | ** | — | |
| NbSe$_2$ | Crystal, Irr: 1.4 GeV Pb (CDs, 3T) | 4.9x10$^{-6}$ | 0.00395 | 0.00626 | |
| **A15, MgB$_2$, BKBO** | | | | | |
| V$_3$Si | Crystal | 2.8x10$^{-6}$ | — | 0.0146 | 38 |
| MgB$_2$ | Film | 2.1x10$^{-7}$ | 0.00178 | — | 39 |
| MgB$_2$ | Film | 4.0x10$^{-5}$ | — | 0.0023 | 38 |
| (K,Ba)BiO$_3$ | Crystal | 3.8x10$^{-4}$ | 0.0128 | 0.0126 | 66 |
| **Borocarbides** | | | | | |
| LuNi$_2$B$_2$C | Crystal | 9.7x10$^{-6}$ | 0.0097 | — | |
| **Fe-SCs: 122 P-doped** | | | | | |
| Ba(As$_{0.67}$P$_{0.33}$)$_2$Fe$_2$ | Film | 1.2x10$^{-4}$ | 0.00485 | 0.0102 | |
| Ba(As$_{0.67}$P$_{0.33}$)$_2$Fe$_2$ | Film, 3% BaZrO$_3$ NPs | | 0.00714 | 0.00847 | |
| **Fe-SCs: 122 Na-doped** | | | | | |
| Ca$_{0.5}$Na$_{0.5}$Fe$_2$As$_2$ | Crystal | 1.4x10$^{-5}$ | 0.00989 | 0.00984 | 8 |
| Ca$_{0.25}$Na$_{0.75}$Fe$_2$As$_2$ | Crystal | 2.9x10$^{-5}$ | 0.03315 | 0.0327 | 8 |
| Ca$_{0.6}$Na$_{0.4}$Fe$_2$As$_2$ | Crystal | 3.3x10$^{-4}$ | — | 0.0349 | 7 |
| Ca$_{0.89}$Na$_{0.11}$FFeAs | Crystal | 2.7x10$^{-3}$ | — | 0.0343 | 7 |
| **Fe-SCs: 122 Co-doped** | | | | | |
| Ba(Fe$_{0.92}$Co$_{0.08}$)$_2$As$_2$ | Crystal$^{††}$ | 4.6x10$^{-4}$ | 0.02278 | 0.024 | 23‡‡ |
| Ba(Fe$_{0.92}$Co$_{0.08}$)$_2$As$_2$ | Crystal$^{††}$, Irr: 1.4 GeV Pb (1 T) | 4.5x10$^{-4}$ | 0.0115 | 0.0111 | 23‡‡ |
| Ba(Fe$_{0.92}$Co$_{0.08}$)$_2$As$_2$ | Crystal$^{††}$, Irr: 1.4 GeV Pb (3.5 T) | 4.3x10$^{-4}$ | 0.0099 | 0.0099 | 23‡‡ |
| Ba(Fe$_{0.925}$Co$_{0.075}$)$_2$As$_2$ | Crystal | 3.2x10$^{-4}$ | — | 0.0265 | 32 |
| Ba(Fe$_{0.925}$Co$_{0.075}$)$_2$As$_2$ | Crystal, Irr: 3 MeV H+, 1x10$^{16}$/cm$^2$ | 3.0x10$^{-4}$ | — | 0.0164 | 32 |
| Ba(Fe$_{0.925}$Co$_{0.075}$)$_2$As$_2$ | Crystal, Irr: 3 MeV H+, 2x10$^{16}$/cm$^2$ | 2.7x10$^{-4}$ | — | 0.0152 | 32 |
| **Fe-SCs: Other 122** | | | | | |
| BaFe$_{1.3}$Ru$_{0.7}$As$_2$ | Crystal | 5.4x10$^{-5}$ | 0.0136 | 0.0194 | 6 |
| Ba$_{0.6}$K$_{0.4}$Fe$_2$As$_2$ | Crystal | 7.2x10$^{-4}$ | — | 0.0389 | 42 |
| Ba$_{0.6}$K$_{0.4}$Fe$_2$As$_2$ | Crystal, Irr: 3 MeV H+, 5.8x10$^{16}$/cm$^2$ | 6.0x10$^{-4}$ | — | 0.0135 | 42 |
| **Fe-SCs: Misc** | | | | | |
| FeSe | Crystal | 1.5x10$^{-4}$ | 0.02497 | 0.025 | 19 |
| FeSe | Crystal, Irr: 3 MeV H+, 5x10$^{16}$/cm$^2$ | 1.5x10$^{-4}$ | 0.01237 | 0.01198 | 19 |
| FeTe$_{0.6}$Se$_{0.4}$ | Crystal | 8.8x10$^{-3}$ | — | 0.0295 | 5 |
| LiFeAs | Crystal | 1.6x10$^{-4}$ | — | 0.0104 | 44 |
| Rb$_{0.8}$Fe$_{1.6}$SeS | Crystal | Unknown | 0.01338 | 0.0244 | |
| **LSCO** | | | | | |
| La$_{1.85}$Sr$_{0.15}$CuO$_4$ | Crystal | 9.1x10$^{-3}$ | 0.04655 | — | |

---

** $S \sim 0.005$ at $\mu_0 H = 200$ Oe (and is faster at 0.3 T than in the irradiated crystal)

†† Annealed

‡‡ Note that all samples in this study were Ba(Fe$_{0.92}$Co$_{0.08}$)$_2$As$_2$, and that the referenced paper includes a few typos, specifying the incorrect doping.

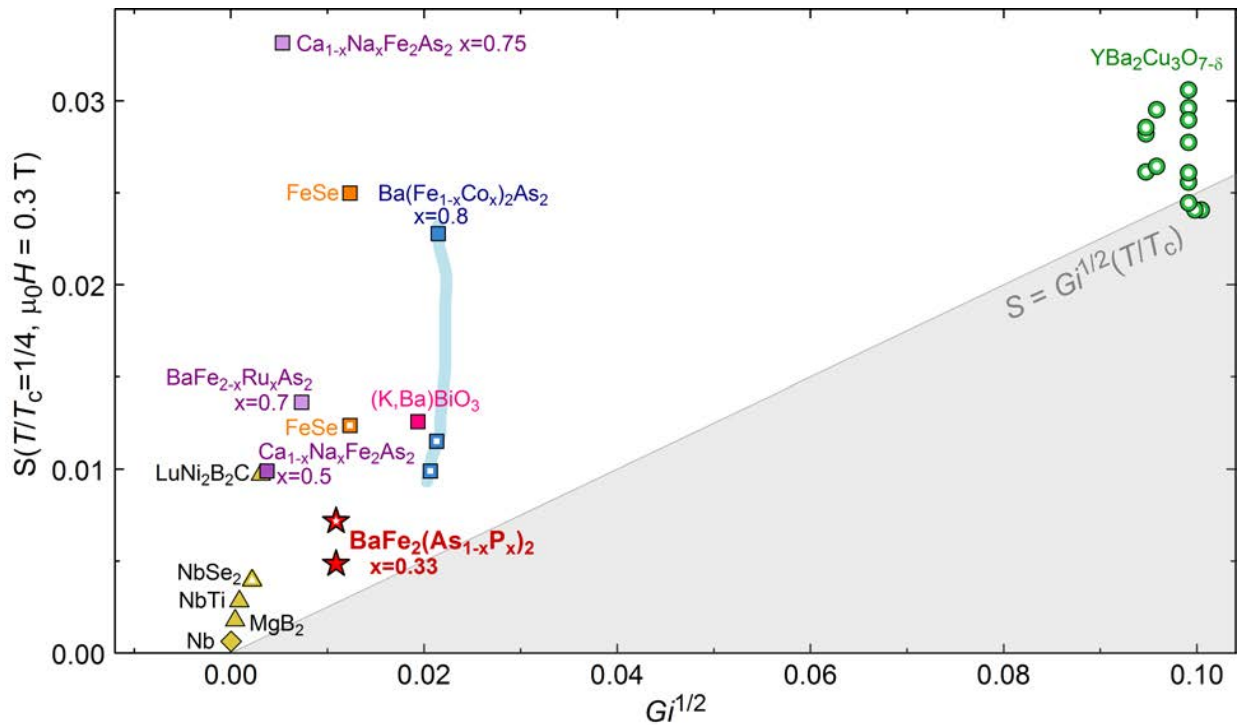

**Supplementary Figure S1. The universal lower limit for creep. (A)** Creep at reduced temperature $T/T_c=1/4$ and field of $\mu_0 H = 0.3$ T for different superconductors plotted versus $Gi^{1/2}$. The open symbols indicate materials for which the microstructure has been modified by either irradiation or incorporation of inclusions. The solid gray line represents the limit set by $Gi^{1/2}T/T_c = Gi^{1/2}/4$.